\documentclass[a4paper,twoside,10pt]{article}

\input{jgrg20.sty}
\usepackage{subfigure}

\begin{document}
%
\pagestyle{fancy}
\fancyhead{}
  \fancyhead[RO,LE]{\thepage}
  \fancyhead[LO]{M. Nakashima}          
  \fancyhead[RE]{CMB Polarization in Einstein-Aether Theory}    
\rfoot{}
\cfoot{}
\lfoot{}
\label{P48}                             
\title{%
  CMB Polarization in Einstein-Aether Theory\footnote{Contribution to the proceedings
  of the conference ``The 20th workshop on General Relativity and Gravitation in Japan (JGRG20).''
  Complete details will be presented in a longer paper~\cite{P48_N10}.}
}
%
\author{%
  Masahiro Nakashima\footnote{Email address: nakashima@resceu.s.u-tokyo.ac.jp}$^{(a,b)}$
  and
  Tsutomu Kobayashi\footnote{Email address: tsutomu@resceu.s.u-tokyo.ac.jp}$^{(b)}$
}
%
\address{%
  $^{(a)}$Department of Physics, Graduate School of Science, The University of Tokyo, Tokyo 113-0033, Japan\\
  $^{(b)}$Research Center for the Early Universe (RESCEU), Graduate School of Science, The University of Tokyo, Tokyo 113-0033, Japan}
%
\abstract{
We study the impact of modifying the vector sector of gravity on the CMB polarization.
We employ the Einstein-aether theory as a concrete example.
The Einstein-aether theory admits dynamical vector perturbations
generated during inflation, leaving imprints on the CMB polarization.
We derive the perturbation equations of the aether vector field
in covariant formalism and compute the CMB B-mode polarization using the modified CAMB code.
It is found that the amplitude of the B-mode signal from the aether field can surpass the one from the inflationary gravitational waves. 
The shape of the spectrum is clearly understood in an analytic way using the tight coupling approximation. 
}

Motivated mainly by the mystery of
the dark components of the Universe,
modification of gravity law from standard general relativity (GR) has been explored much.
It is commonly made by adding an extra scalar degree of freedom as in
$f(R)$ theories. It is also possible to modify the spin-2 sector
as in massive gravity and bi-gravity theories.
In this paper, we are going to consider
a hypothetical {\em vector} degree of freedom of gravity.
Specifically, we shall focus on the Einstein-aether (EA) theory
proposed by Jacobson and Mattingly~\cite{P48_J01},
in which a fixed norm vector field
with a Lorentz-violating vacuum expectation value
takes part in the gravitational interaction.

The purpose of the present paper is to clarify the impact of
the aether vector field on the CMB polarization.
The CMB polarization arises from all the three types of cosmological perturbations, i.e.,
scalar, vector, and tensor perturbations.
Among them, the vector perturbations most effectively generate the B-mode polarization~\cite{P48_H97}, though 
the effect has been less investigated
because the vector mode decays unless sourced, {\em e.g.}, by topological defects or the neutrino anisotropic stress~\cite{P48_L04}.
Modifying the vector sector of gravity would add
yet another possibility of producing vector perturbations,
leaving a unique signature in the CMB polarization due to nontrivial dynamics of the aether field.



The action of the EA theory is given by
\begin{eqnarray}
{\cal S}=\frac{1}{2\kappa}\int d^4x\sqrt{-g}[{\cal R}-c_1\nabla_aA^b\nabla^aA_b-c_2(\nabla_bA^b)^2 - c_3\nabla_aA^b\nabla_bA^a \nonumber \\
-c_4A^aA^b\nabla_aA^c\nabla_bA_c]+\lambda(A_bA^b-1)]+{\cal S}_m,
\end{eqnarray}
where $\kappa=8\pi G$, ${\cal R}$ is the Ricci scalar, $A_{a}$ is the aether field, and ${\cal S}_m$ is the action of ordinary matter.
Variation with respect to $A^a$ yields the equation of motion for the aether
and variation with respect to the Lagrange multiplier $\lambda$
gives the fixed norm constraint, $A_aA^a=1$.
In the rest of the paper we
use the following abbreviations: $c_{13}=c_1+c_3, c_{14}=c_1+c_4, \alpha =c_1+3c_2+c_3$.

To describe background cosmology and the evolution of vector perturbations,
we employ the covariant equations obtained by the method of $3+1$ decomposition.
We begin with splitting physical quantities with respect to observer's 4-velocity $u^a$.
Following the usual procedure, the projection tensor is defined as
$h_{ab}:=g_{ab}-u_au_b$. We define time derivative as $\dot{T}^{a\cdots}_{b\cdots} := u^{c}\nabla_{c}T^{a\cdots}_{b\cdots}$ and covariant spatial derivative as $D^aT^{b\cdots }_{c\cdots } := h^a_ih^b_j\cdots h^k_c\cdots \nabla^i T^{j\cdots }_{k\cdots }$.
The energy-momentum tensor for each matter component
and $\nabla_{a}u_{b}$ are expressed respectively as
\begin{equation}
T_{ab}^{(i)} = \rho^{(i)} u_a u_b- p^{(i)} h_{ab}+2 q_{(a}^{(i)}u_{b)}+\pi_{ab}^{(i)},
\ \ \nabla_a u_b = \frac{1}{3}\theta h_{ab}+\sigma_{ab}+\omega_{ab}-u_a\dot u_b.
\end{equation}
The energy-momentum tensor for the aether, $T^{(A)}_{ab}$,
is also written in the same form as above.
The expansion $\theta$ may be written as $\theta=3\dot S/S$, where $S$ is the averaged scale factor.

At zeroth order, $A^a=u^a$, so that 
the energy density and the pressure of the aether are given respectively by
$\rho^{(A)} = \alpha\theta^2/6$ and $p^{(A)}=-\alpha(2\dot\theta+\theta^2)/6$.
The Friedman equation is thus given by 
$3{\cal H}^2 = \kappa S^{2}\rho/(1-\alpha/2)$
where $\rho$ is the total energy density of ordinary matter
and we have introduced the conformal Hubble parameter, ${\cal H}:=S\theta/3$.
The background effect of the aether is just to rescale the gravitational constant $\kappa$.

Let us move on to the dynamics of vector perturbations.
We choose $u_a$ to be hypersurface orthogonal,
so that $\dot u_b=0$ at linear order.
At this order, the aether field can be written as 
$A_b = u_b+D_b V^{(s)} + V_b$,
where $V^{(s)}$ corresponds to a scalar perturbation which we do not consider in this paper,
while $V_b$ a vector perturbation that satisfies $D_bV^b=0$.
Each perturbation variable can be expanded using
the transverse eigenfunctions $Q_{a\cdots}^{\pm}$:
$V_{a}=\sum VQ_{a}^{\pm}$, $q_{a}^{(i)}=\sum q^{(i)} Q_{a}^{\pm}$,
$\sigma_{ab}=\sum (k/S) \sigma Q_{ab}^{\pm}$, and $\pi_{ab}^{(i)}=\sum\Pi^{(i)}Q_{ab}^\pm$,
where $k$ is the eigenvalue.
It is convenient to write the relevant equations
in terms of the coefficient functions $V$, $q^{(i)}$, ...\,.


From the equation of motion for the aether, we obtain the evolution equation for $V$:
\begin{equation}
c_{14}\left[V''+2{\cal H}V'+\left({\cal H}^2+{\cal H}'\right)V\right]
+\alpha\left({\cal H}^2-{\cal H}'\right)V+c_1k^2 V
=-\frac{c_{13}}{2}k^2\sigma.\label{ae-evolution}
\end{equation}
This shows that the fluctuation of the aether obeys
the wave equation which is similar to the evolution equation for cosmological tensor perturbations.
The crucial difference is the effective mass term which is dependent on
the expansion rate ${\cal H}$ and the model parameters.
The fluctuation of the aether leads to
the effective heat-flux vector
$q_a^{(A)} =-c_{13}D^b [\sigma_{ab}+D_{(a}V_{b)} ]$, which
gives rise to the additional contribution to the momentum constraint equation.
The momentum constraint under the influence of the aether is thus given by
\begin{equation}
k^2\sigma=\frac{1}{1+c_{13}}\left(2\kappa S^2 \sum_{i} q^{(i)} - c_{13}k^2V\right).
\label{mom-const}
\end{equation}
Here, $q^{(i)}$ is determined through the individual matter equations.
In calculating the CMB power spectrum
we simultaneously solve the equations of motion
for other ordinary matter and
the multipole moment equations for photons and neutrinos,
as well as the equations derived above.

We now fix the initial condition for each variable at the early radiation-dominant epoch. 
This is done by a series expansion in terms of the conformal time $\eta$,
following \cite{P48_L04}, but now taking into account the presence of
the aether.
Neglecting the ${\cal O}(k^2)$ terms, the perturbed equation of motion for the aether can be solved to give
\begin{eqnarray}
V={\cal A}_k\eta^\nu\left[1+{\cal O}(\eta)\right],
\quad
\nu :=\frac{-1+\sqrt{1-8\alpha/c_{14}}}{2},
\end{eqnarray}
implying
that $V$ can grow on superhorizon scales.
Here, we have dropped the decaying mode solution
which is not regular at $\eta\to 0$.
The coefficient ${\cal A}_k$ is determined from the primordial spectrum of
the aether fluctuation~\cite{P48_Ar10}.
Requiring that scalar isocurvature modes do not grow,
we consider the range $0\le \nu\le 1$~\cite{P48_Ar10}.
As for the other variables, the appropriate early time solutions are found to be
\begin{eqnarray}
\sigma =-\frac{\nu^*}{\nu^*+4R^*}\frac{c_{13}}{1+c_{13}}{\cal A}_k\eta^\nu,\ \
q^{(\gamma)}=q^{(b)}=q^{(\nu)}=0, \ \
\frac{\Pi^{(\nu)}}{\rho^{(\nu)}}=-\frac{8}{15(1+\nu)}\frac{\nu^*}{\nu^*+4R^*}
\frac{c_{13}}{1+c_{13}}{\cal A}_kk\eta^{1+\nu},
\end{eqnarray}
where we defined
$R^*:=[(1-\alpha/2)/(1+c_{13})]\rho^{(\nu)}/(\rho^{(\gamma)} +\rho^{(\nu)})$ and
$\nu^*:=(5/2)(1+\nu)(2+\nu)$, with the superscripts $\gamma$, $b$, and $\nu$
denoting photons, baryons, and neutrions, respectively.

In the AE theory, primordial vector perturbations are generated quantum mechanically
during inflation.
Once the inflation model and the reheating history are specified,
one can determine the primordial spectrum of the vector perturbation
and hence ${\cal A}_k$, as discussed in~\cite{P48_Ar10}.
We separate the issue of the perturbation evolution during inflation
from the subsequent evolution, and set simply
${\cal A}_k={\cal A}_0 k^{(n_v-3)/2}$
({\em i.e.}, the primordial spectrum ${\cal P}_V\propto k^{n_v}$),
where ${\cal A}_0$ is a constant.


\begin{figure}[t]
\centering
\includegraphics[keepaspectratio=true,width=9cm]{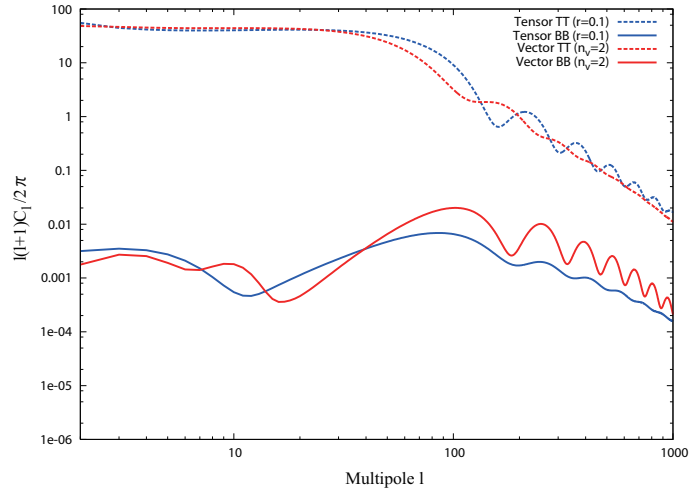}
\caption{
CMB B-mode polarization and temperature anisotropy power spectra in the EA theory. For comparison, those from the tensor perturbation in standard general relativity are also plotted in the case of the tensor-to-scalar ratio $r=0.1$. In this figure, $c_{1}=-0.2,c_{13}=-0.3,c_{14}=-\alpha=-0.2$, and dimensionless primordial power spectra are $\mathcal{P}_{V}\propto k^{n_{v}}$ and $\mathcal{P}_{T}\propto k^{0}$.
}
\label{fig:P48_ex1}
\end{figure}

We have completed the numerical calculation using all the ingredients derived
above and the CAMB code~\cite{P48_L00} modified so as to incorporate the presence of the aether. An example of our numerical results is presented in the Fig.~\ref{fig:P48_ex1}.
For comparison, we show
contributions from inflationary gravitational waves in GR, assuming that
the tensor-to-scalar ratio is given by $r=0.1$.
The amplitude ${\cal A}_0$ is adjusted so that the low-$\ell$ TT spectrum from the vector perturbation
has the same magnitude
as this primordial tensor contribution.
We see in this case that the BB spectrum in the EA theory
is larger than that from primordial tensor modes at $\ell\gtrsim 100$,
and hence the B-mode is potentially detectable in future CMB observations
aiming to detect $r={\cal O}(0.1)-{\cal O}(0.01)$.




Let us try to understand the shape of the B-mode angular power spectrum $C_{\ell}^{BB}$
in the EA theory in an analytic way.
We start with the integral solution for the moment $B_\ell^V$
 of the B-mode polarization in the covariant formalism:
\begin{equation}
B_{\ell}^{V}(\eta_{0}) = -\frac{\ell-1}{\ell+1}\int^{\eta_{0}} d\eta \dot{\tau} e^{-\tau} \Psi_{\ell}(x)\zeta,
\quad x = k(\eta_0-\eta),
\end{equation}
with $\Psi_{\ell}(x)= \ell j_{\ell}(x)/x$ and
$\zeta = (3/4)I_{2}-(9/2)E_{2}$. Here,
$j_{\ell}(x)$ is a spherical Bessel function, $\tau$ is the optical depth, $I_{\ell}$ is the
angular moment of the fractional photon density distribution, and $E_{\ell}$ is the
moment of the E-mode polarization. Using the tight coupling approximation, we obtain 
\begin{equation}
\zeta \simeq \frac{4k}{15\dot{\tau}}\sigma \simeq -\frac{4k}{15\dot{\tau}}\frac{c_{13}}{1+c_{13}}V
\end{equation}
It turns out that ignoring $q^{(i)}$ in Eq.~(\ref{mom-const})
is a good approximation. We thus arrive at~\cite{P48_Ar10}
\begin{equation}
V^{\prime\prime}+2{\cal H}V^{\prime}+c_{v}^{2}k^{2}V+\left[\left(1+\frac{\alpha}{c_{14}}\right){\cal H}^{2}-\left(1-\frac{\alpha}{c_{14}}\right){\cal H}^{\prime}\right]V\simeq 0,\quad
c_{v}^{2}=\frac{c_{1}}{c_{14}}\left[1-\frac{c_{13}^{2}}{2c_{1}(1+c_{13})}\right].
\end{equation}
On superhorizon scales we find
$V\propto S^\nu$ in the radiation-dominant stage, as already derived, and
$V\propto S^{{\nu_m}/2}$ with $\nu_m = (-3+\sqrt{1-24\alpha/c_{14}})/2$ in the matter-dominant stage.
On subhorizon scale, $V$ simply decays similarly to tensor perturbations, $V\propto S^{-1}$.
These relations can be mapped into the wavenumber dependence of $V$ at recombination ($\eta=\eta_{\rm rec}$) as
$V \propto {\cal A}_k\ (k<1/c_{v}\eta_{\rm rec})$, $V \propto k^{-2-\nu_m}{\cal A}_k \ (1/c_{v}\eta_{\rm rec}<k<1/c_{v}\eta_{\rm eq})$, and $V \propto k^{-1-\nu}{\cal A}_k\ (1/c_{v}\eta_{\rm eq}<k)$,
where $\eta_{\rm eq}$ refers to the radiation-matter equality time.

\begin{figure}[t]
\subfigure[]{%
\includegraphics[keepaspectratio=true,width=8cm]{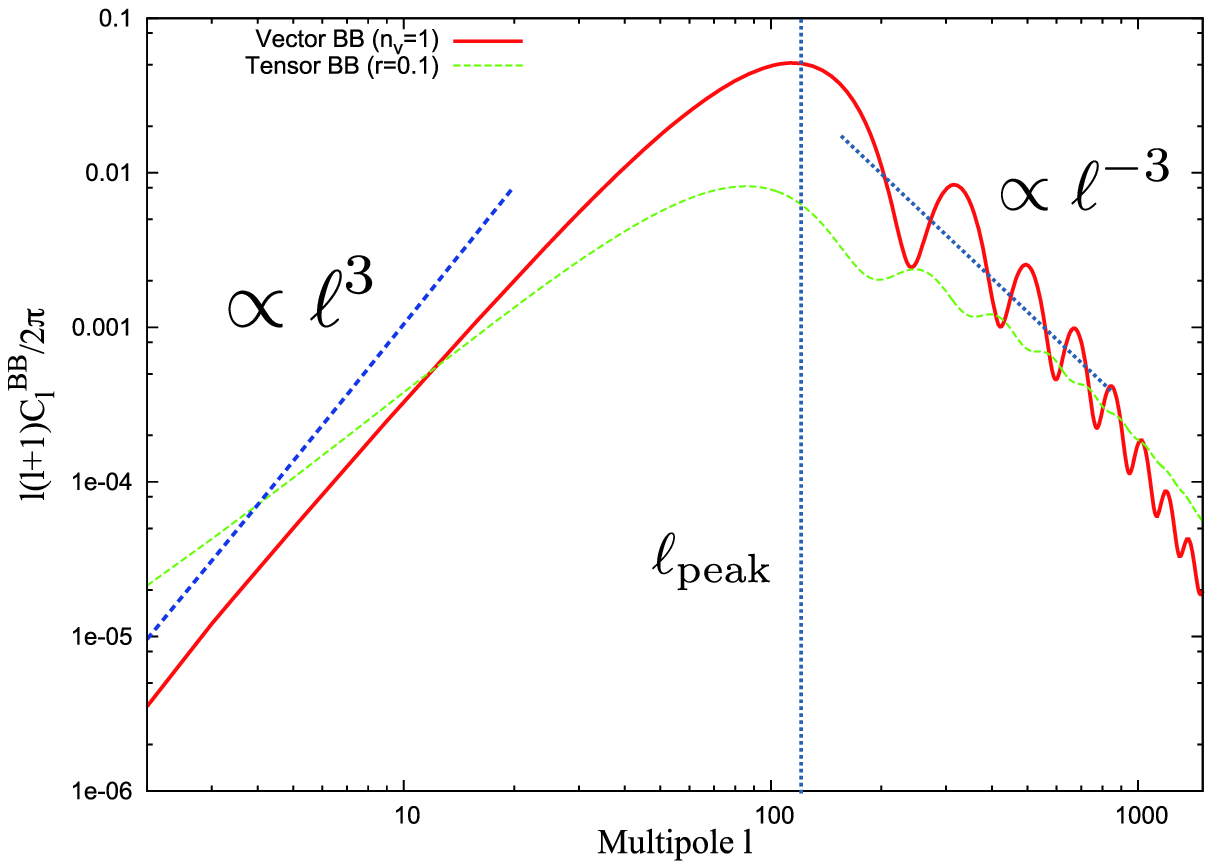}}%
\subfigure[]{%
\includegraphics[keepaspectratio=true,width=8cm]{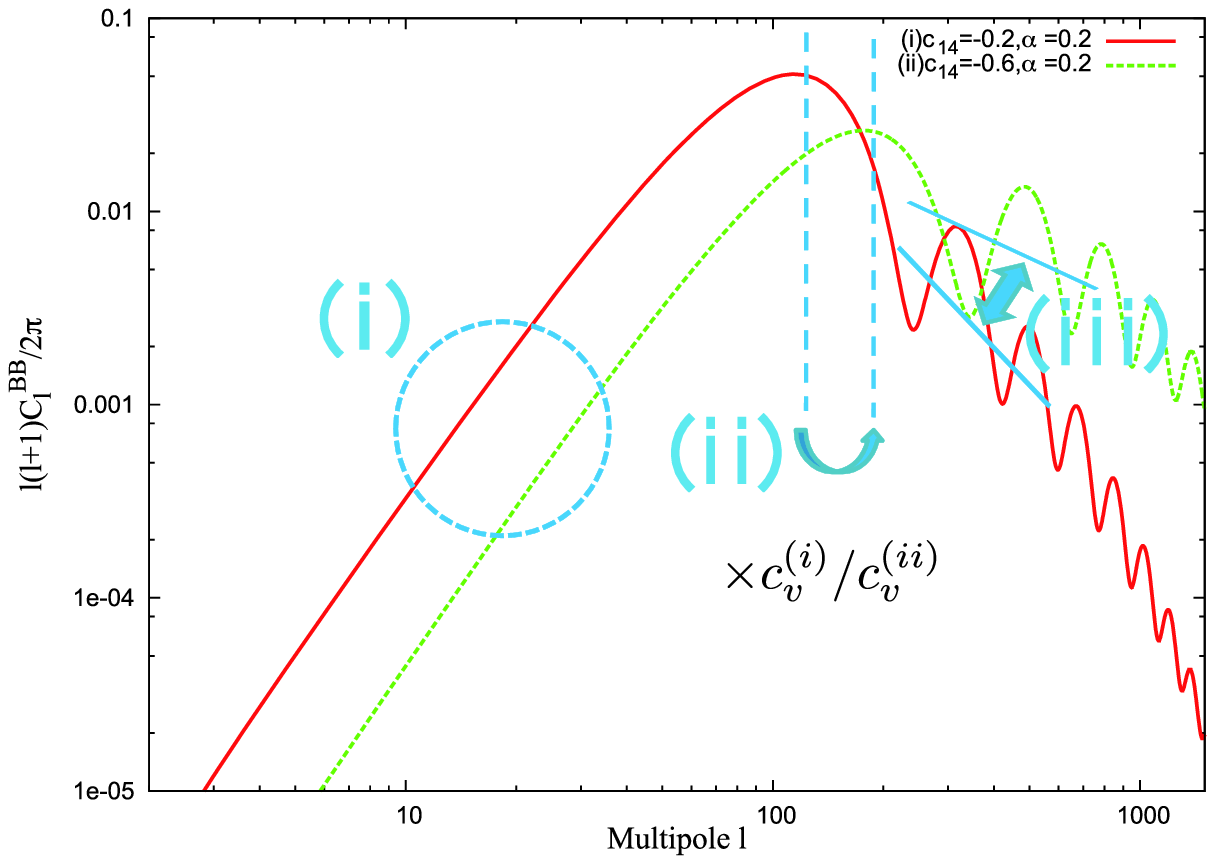}}%
\caption{(a) Scaling for the illustrative case with $n_v=1$ and $\alpha=-c_{14}=0.2$.
The other parameters are given by $c_{13}=-0.3$, and $c_1=-0.1$;
(b) Parameter dependence of the spectrum.
In the two examples $c_{14}$ is different while the other parameters
are fixed as $n_v=1$, $c_{13}=-0.3$, and $c_1=-0.1$. The primordial amplitudes are arbitrary.}
\label{fig:P48_ex2}
\end{figure}

The CMB B-mode power spectrum is roughly expressed as
$C_{\ell}^{BB}\sim \int {\cal P}_{V}(k) B_{\ell}^{V}B_{\ell}^{V} d\ln k$.
Using the approximation $\dot{\tau}e^{-\tau}\simeq \delta(\eta-\eta_{\rm rec})$,
$B_\ell^V(\eta_0)$ can be written as
$B_{\ell}^{V}(\eta_{0})\simeq (k/\dot{\tau})[ c_{13}/(1+c_{13})]
V(\eta_{\rm rec})\Psi_{\ell}[k(\eta_{0}-\eta_{\rm rec})]$. ($\eta_0$ is the present time.) 
Since the projection factor $\Psi_{\ell}(x)$ has a peak at $\ell \simeq x$,
the angular power spectrum reduces approximately to
\begin{equation}
C_{\ell}^{BB} \sim \left(\frac{V}{{\cal A}_k}\right)^{2}_{k=\ell/(\eta_{0}-\eta_{\rm rec})}
\int k^{2+n_v}[\Psi_{\ell}(k(\eta_{0}-\eta_{\rm rec}))]^{2} d\ln k.
\end{equation}
For example,
for $n_v=1$ and $c_{14}=-\alpha$,
the above integral 
can be evaluated to give 
the scaling $\ell(\ell+1)C_{\ell}^{BB}\propto\ell^{3}\ (\ell < \ell_{\rm rec}:=(\eta_0-\eta_{\rm rec})/c_{v}\eta_{\rm rec})$,
$\ell^{-3}\ (\ell_{\rm rec}< \ell < \ell_{\rm eq}:=(\eta_0-\eta_{\rm eq})/c_{v}\eta_{\rm eq})$, and
$\ell^{-1}\ (\ell_{\rm eq}<\ell)$, showing a peak at $\ell_{\rm peak}\sim\ell_{\rm rec}$.
This behavior can indeed be seen in Fig.~\ref{fig:P48_ex2}(a), though
the scaling at $\ell>\ell_{\rm eq}$ is hidden by the other effect and hence is not obvious.
(Here, for comparison, the B-mode spectrum from tensor perturbations in standard GR are also plotted.)

We can also gain an understanding of
how the shape of the angular power spectrum depends on
the model parameters.
From Fig.~\ref{fig:P48_ex2}(b)
one can confirm the following three things:
(i) since the angular power spectrum on the largest scales $\ell < \ell_{\rm rec}$
depends only on the primordial spectrum, the plotted examples show the same scaling;
(ii) the peak position is inversely proportional to the sound velocity
of the aether vector perturbation $c_{v}$;
(iii) the difference of the small scale scaling arises due to the difference of the
growth rate of $V$ on superhorizon scales.
The detailed discussion on the analytic estimate will be provided in~\cite{P48_N10}.

\end{document}